\documentclass[twocolumn,english,pra,showpacs,preprintnumbers,superscriptaddress]{revtex4-1}
\usepackage[utf8]{inputenc}
\usepackage{color}
\usepackage{amsmath,amssymb}
\usepackage{amssymb}
\usepackage{graphicx}
\usepackage{textcomp}
\usepackage{dcolumn}
\usepackage{bm}
\usepackage[right]{eurosym}
\usepackage{float}
\usepackage[english]{babel}
\usepackage{blindtext}
\usepackage{babel}
\usepackage[normalem]{ulem}

\begin{document}

\title{Photoelectron spectroscopy of coronene molecules embedded in helium nanodroplets}


\author{L. Ben Ltaief}
\affiliation{Department of Physics and Astronomy, Aarhus University, 8000 Aarhus C, Denmark}
\author{M. Shcherbinin}
\affiliation{Indian Institute of Science Education and Research, Pune 411008, India}
\author{S. Mandal}
\affiliation{Indian Institute of Science Education and Research, Pune 411008, India}
\author{S. R. Krishnan}
\affiliation{Department of Physics, Indian Institute of Technology, Madras, Chennai 600 036, India}

\author{R. Richter}
\affiliation{Elettra-Sincrotrone Trieste, 34149 Basovizza, Trieste, Italy}
\author{S. Turchini}
\affiliation{Istituto Struttura della Materia-CNR (ISM-CNR), 00133 Roma, Italy}
\author{N. Zema}
\affiliation{Istituto Struttura della Materia-CNR (ISM-CNR), 00133 Roma, Italy}
\author{M. Mudrich}
\email{mudrich@phys.au.dk}
\affiliation{Department of Physics and Astronomy, Aarhus University, 8000 Aarhus C, Denmark}
\affiliation{Department of Physics, Indian Institute of Technology, Madras, Chennai 600 036, India}

\begin{abstract}
We present the first measurement of a one-photon extreme-ultraviolet photoelectron spectrum (PES) of molecules embedded in superfluid helium nanodroplets. The PES of coronene is compared to gas phase and the solid phase PES, and to electron spectra of embedded coronene generated by charge transfer and Penning ionization through ionized or excited helium. The resemblence of the He-droplet PES to the one of the solid phase indicates that mostly Cor clusters are photoionized. In contrast, the He-droplet Penning-ionization electron spectrum is nearly structureless, indicating strong perturbation of the ionization process by the He droplet. These results pave the way to extreme ultraviolet photoelectron spectroscopy (UPS) of clusters and molecular complexes embedded in helium nanodroplets.

\end{abstract}

\date{\today}

\maketitle

\section{Introduction}
Over the past decades, helium (He) nanodroplets have established themselves as flying nano-cryo-laboratories to investigate cold molecular reactions~\cite{farnik2005ion,mauracher2018cold,henning2019experimental}, novel nanostructures~\cite{gomez2012traces,boatwright2013helium,latimer2014preparation,haberfehlner2015formation,wu2016development}, and high-resolution molecular spectra~\cite{callegari2001helium,Toennies:2004,Stienkemeier:2006,Mudrich:2014}. The main benefits of He nanodroplets are that individual molecules or small clusters of atoms or molecules can be isolated in an ultracold (0.37~K), highly dissipative but weakly perturbing environment (He nanodroplet isolation, HENDI). High-resolution absorption and emission spectra of the embedded species can be recorded in the infrared, visible, up to ultraviolet spectral regions.

Extending HENDI spectroscopy to the extreme-ultraviolet and x-ray range to probe valence and inner-shell spectra of embedded (``dopant'') molecules is more challenging, though. When the dopants or the He droplets are photoionized, strong interactions of the produced photoions and electrons with the local environment tend to massively shift and broaden spectral lines and to alter the fragmentation dynamics compared to the gas phase~\cite{Mudrich:2014}. Moreover, whenever the photon energy $h\nu$ is high enough to even excite or ionize the He droplets, \textit{i.~e.} $h\nu > 21~$eV, indirect Penning ionization of the dopants through excited He or charge transfer (CT) ionization through ionized He usually predominates over direct dopant photoionization. Therefore, only few photoelectron spectroscopic studies of dopants in He droplets have been reported, all of which employed resonant multi-photon ionization schemes~\cite{Radcliffe:2004,Loginov:2005,Loginov:2007,Loginov:2008,Fechner:2012,Thaler:2018,Dozmorov:2018,kazak2019photoelectron,Thaler:2018}. In resonant multi-photon ionization using nanosecond laser pulses, nuclear rearrangement following excitation occurs on the picosecond time scale, such that photoelectrons are emitted from a transient state of the system. 

Only recently, we have reported the first direct one-photon ionization electron spectra of He droplet-embedded ground state species -- small clusters of xenon (Xe) atoms ionized by soft x-ray synchrotron radiation~\cite{ltaief:2020direct}. In that experiment, electron emission from the Xe dopants was enhanced by tuning the photon energy to the absorption `giant resonance' (4d-shell ionization) around $h\nu =100~$eV. As the absorption cross section of He is much smaller at that photon energies, direct photoionization of the Xe dopants was comparable or even higher in intensity to CT ionization through the photoionized He droplets. 

Here, we report the first measurements of a photoelectron spectrum (PES) of a molecule, the polycyclic aromatic hydrocarbon coronene (Cor), embedded in He nanodroplets. We suppress photoionization of the He droplets and subsequent CT ionization of the dopants by tuning the photon energy below the lowest absorption bands of He droplets, $h\nu <21~$eV. Cor is particularly well-suited for this purpose owing to its high photostability and its large absorption cross section of about 500 Mbarn at $h\nu =18.5~$eV~\cite{khakoo:1990}. The resulting electron spectra are compared to Penning ionization electron spectra, which are more intense owing to the large resonant-absorption cross section of He droplets and a high Penning ionization efficiency~\cite{Wang:2008,Buchta:2013,Shcherbinin:2018,Ltaief:2019,Mandal:2020}. We find that the He-droplet PES resemble those previously measured for solid Cor, indicating that mostly Cor clusters are probed in the He droplets.

\section{Experimental methods}
The experiments are performed using a He nanodroplet apparatus combined with a velocity-map imaging photoelectron-photoion coincidence (VMI-PEPICO) detector at the CiPo beamline of Elettra-Sincrotrone Trieste, Italy. The use of a gold coated normal incidence monochromator (2400 l/mm) ensures a high spectral resolution 
and the suppression of higher-order radiation to a high degree.

The apparatus has been described in detail elsewhere~\cite{Buchta:2013,BuchtaJPC:2013}. Briefly, a beam of He nanodroplets is produced by continuously expanding pressurized He (50~bar) of high purity out of a cold nozzle (14~K) with a diameter of 5~$\mu$m into vacuum, resulting in a mean droplet size of $\bar{N}_\mathrm{He} = 2.3\times 10^4$ He atoms per droplet. Further downstream, the beam passes a mechanical beam chopper used for discriminating droplet-beam correlated signals from the background. The He droplets are doped with Cor molecules by passing through a vapor cell containing elementary Cor crystalline powder heated to 150~$^\circ$C. At this temperature, the He nanodroplets pick up 1 Cor molecule with highest probability, 37\,\%. As the pick-up process usually obeys the Poissonian statistics~\cite{Lewerenz:1995}, fractions of the He droplets pick up 2 molecules (18\,\%) or more (8\,\%). 

In the detector chamber, the He droplet beam crosses the synchrotron beam in the center of the VMI-PEPICO detector at right angles. By detecting either electrons or ions using the VMI detector in coincidence with the corresponding particles of opposite charge with the TOF detector, we obtain either ion mass-correlated electron images or mass-selected ion images. Kinetic energy distributions of electrons and ions are obtained by Abel inversion of the images~\cite{Dick:2014}. The energy resolution of the electron spectra obtained in this way is $\Delta E/E=6$\%.

\begin{figure}[htb]
	\center
	\includegraphics[width=1\columnwidth]{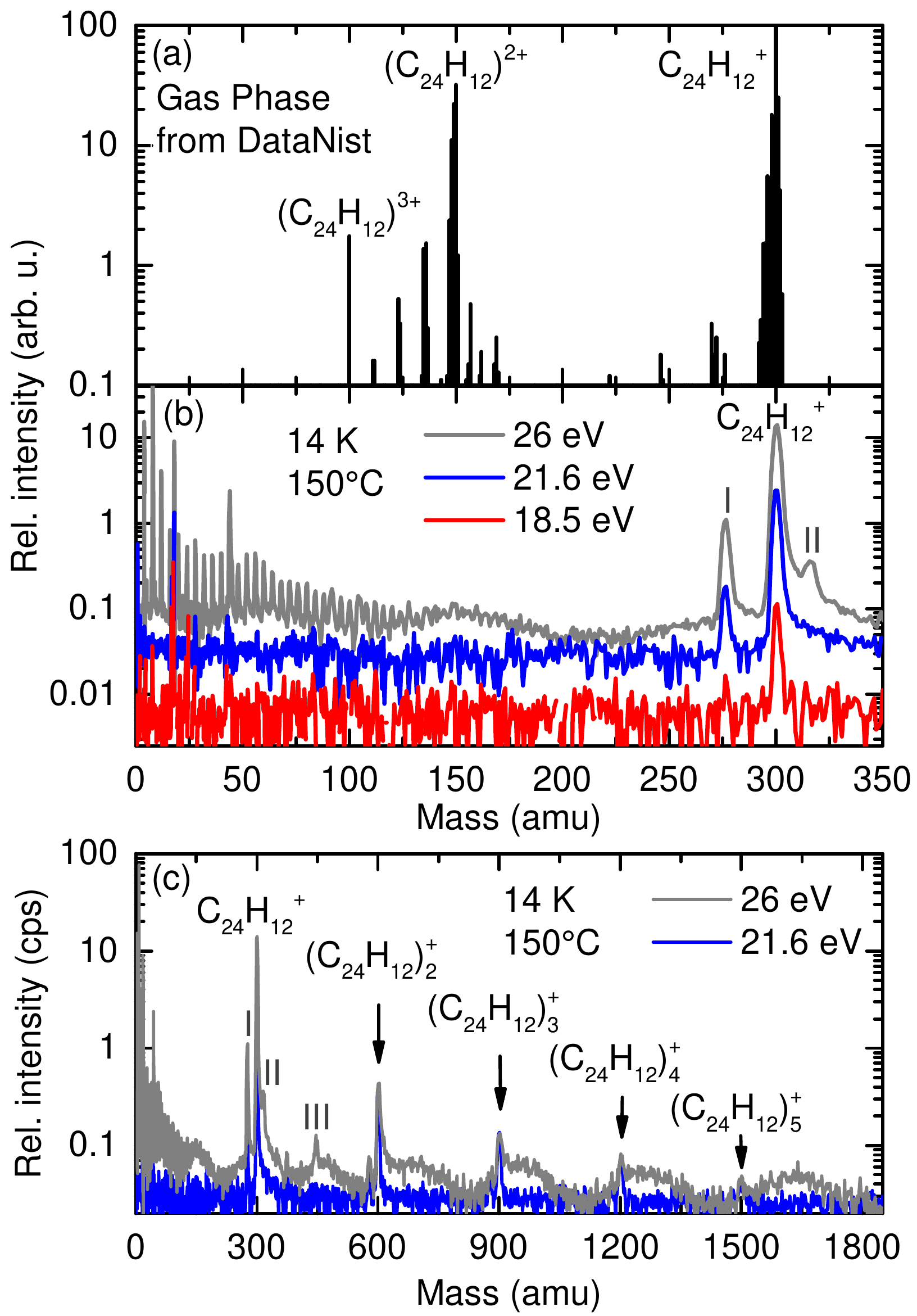}\caption{\label{fig1} (a) Reference mass spectrum of coronene recorded by electron impact ionization~\cite{DataNist}. (b), (c) Mass spectra recorded by photoionization of coronene-doped He nanodroplets at different photon energies.}
\end{figure}
\section{Results and discussion}
Figure 1 b) shows three typical mass spectra recorded for Cor molecules embedded in He nanodroplets at $h\nu =18.5$, $21.6$, and $26~$eV. Contributions from gasphase Cor in the background gas were subtracted. For reference, we show in a) the electron impact ionization mass spectrum of gas phase Cor from the NIST Mass Spectrometry Data Center~\cite{DataNist}. Clearly, the most abundant ion in all three mass spectra is singly charged Cor, C$_{24}$H$_{12}^{+}$, at mass $300$~amu. The mass peak at 276.4~amu (peak I) corresponds to C$_{22}$H$_{12}^{+}$ which is an impurity in the Cor samples. The one at 316.0~amu (peak II) is  C$_{24}$H$_{12}$O$^{+}$ due to oxidization of Cor ions in the reaction with H$_{2}$O molecules which are also present in the He nanodroplets in small amounts. On the larger mass scale, small unfragmented Cor clusters, Cor$_n$ with $n=2$-$5$ are clearly visible, see panel (c). The abundance of the Cor clusters reflects the Poissonian distribution of oligomers formed by doping with multiple Cor molecules~\cite{Lewerenz:1995}. Thus, Cor is very stable against fragmention upon ionization, and mostly unfragmented Cor molecules and Cor$_n$ are detected. 

The formation of Cor$_n$ ions at $h\nu =26~$eV is due to charge transfer ionization through the ionized He nanodroplet~\cite{Callicoatt:1996}, whereas at $h\nu =21.6~$eV, Cor is ionized by Penning ionization as previously reported for other atomic and molecular dopants~\cite{Wang:2008,Buchta:2013,Shcherbinin:2018,Ltaief:2019,Mandal:2020}. At $h\nu =21.6~$eV, He nanodroplets are resonantly excited to their 1s2p$\,^1$P state, and Penning ionization occurs after relaxation to the 1s2s$\,^{1,3}$S state~\cite{Wang:2008,Ltaief:2019,Mudrich:2020}. In the low-mass range (1-130~amu) of the mass spectrum recorded at $h\nu=26~$eV, a series of He$_k^+$ complexes is present with intensities reaching up to that of Cor$^+$. The shoulders extending the Cor$_n^+$ peaks to higher masses are due to charged Cor-He complexes, so-called snowballs. The detailed structures of $[$Cor$_n$He$_k]^+$ snowballs have recently been studied using electron impact ionization mass spectrometry~\cite{kurzthaler2016adsorption,goulart2017structure}. Peak III at mass 446.4~amu nearly matches the mass of the doubly charged Cor trimer cation, $[$C$_{24}$H$_{12}]_{3}^{2+}$, which was previously observed in Ref.~\cite{mahmoodi:2018}. 

The doubly charged Cor cation, present in the electron-impact mass spectrum, shown for reference in panel a), has an appearance energy about $18.7~$eV. In our photoionization mass spectra, no multiply charged Cor is seen because the efficiency of double ionization at energies up to 26~eV is negligible as compared to single-ionization~\cite{denifl:2006}. Note also that in a previous electron-impact study of gasphase Cor, a peak at a mass of 75~amu was observed and assigned to the C$_{6}$H$_{3}^{+}$ fragment ion~\cite{denifl:2006}. Possibly, fragmentation is suppressed due to the cold environment provided by the He nanodroplets as it was previously observed~\cite{lewis:2004}. 

Indirect ionization of molecules doped in He nanodroplets by charge transfer or Penning ionization through ionized or excited He nanodroplets, respectively, has been reported multiple times. Here we report for the first time direct ionization of molecules embedded in He nanodroplets by a one-photon process. The mass spectrum recorded at $h\nu =18.5$~eV, well below the lowest absorption resonances of He nanodroplets~\cite{Joppien:1993}, shows a clear peak at the mass of Cor and of Cor clusters up to the trimer (not shown in figure 1 (c)). Previous attempts to record PES of dopants in He droplets using synchrotron radiation were hampered by the presence of higher-order radiation which is hard to suppress. This energetic radiation efficiently photoionizes the He droplets owing to their large absorption cross section given by that of the He atom times the number of He atoms in the droplet, typically ranging between $10^3$ and $10^5$. Once ionized, the He droplet transfers its charge to the dopant with high efficiency~\cite{Callicoatt:1996}, thereby creating dopant ions in coincidence with an electron emitted by He. 

However, in the present experiment the higher-order content is low owing to the use of a normal incidence monochromator. Our claim that at $h\nu =18.5$~eV, Cor$^+$ is predominantly formed by direct photoionization, is supported by the fact that the mass spectrum significantly differs from the one measured in the regime of He droplet  photoionization at $h\nu=26$~eV: The series of He$_k^+$ mass peaks is entirely missing. However, the count rate of Cor$^+$ is lower compared to the measurement at $h\nu=26$~eV by about 2 orders of magnitude. This factor roughly matches the reduction of the absorption cross section of Cor (500~Mbarn) compared to He droplets ($23.000\times 6.8\,\mathrm{Mbarn}=156$~Gbarn)~\cite{Samson:2002}. 

\begin{figure}[htb]
	\center
	\includegraphics[width=1\columnwidth]{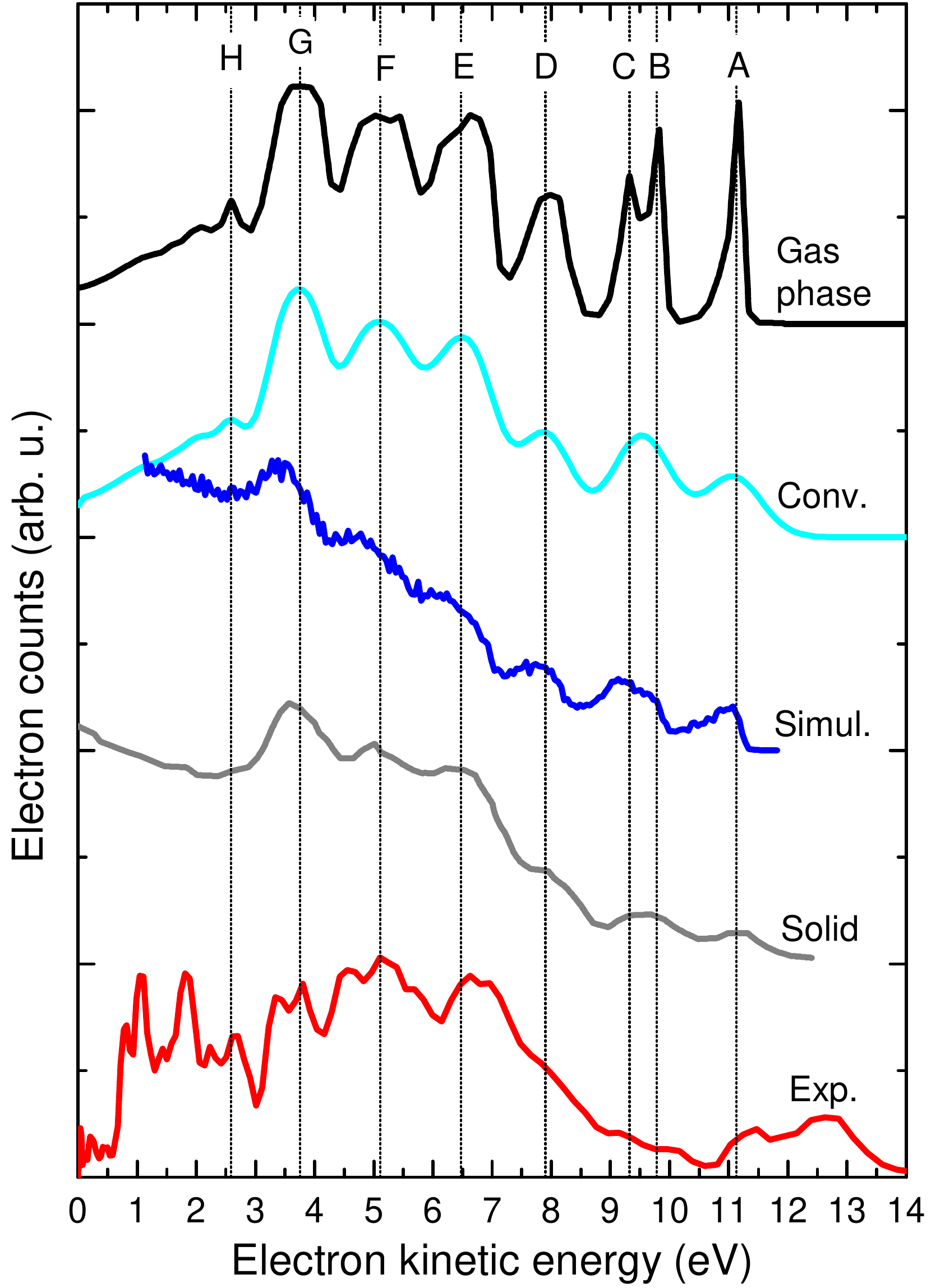}\caption{\label{fig2} He nanodroplet-correlated photoelectron spectrum of coronene (red) recorded at $h\nu = 18.5~$eV in comparison with previously measured coronene gas phase (black)~\cite{acocella:2016} and solid phase (green)~\cite{deleuze:2004,yamakado:1998} photoelectron spectra. The cyan curve shows the gas phase spectrum convoluted with the instrument function. The blue curve is a simulated spectrum based on the gas-phase photoelectron spectrum taking electron–He elastic scattering into account for a He droplet size of about $2\times 10^4$ He atoms per droplet.}
\end{figure}	

Let us now inspect the electron spectrum recorded at $h\nu =18.5~$eV for doped He droplets in coincidence with Cor ions, see the red line in Figure 2. This photon energy is near the maximum of the absorption cross section of Cor and sufficiently far below the lowest absorption bands of He nanodroplets (21~eV) to avoid their excitation or ionization. The peak around 12.4~eV mainly stems from photoelectrons emitted by He droplets due to the residual second-order radiation ($2\times h\nu =37~$eV) still present to a small extent. The area of this peak amounts to $\lesssim 6$\% of the total coincident electron counts, which confirms that the electron spectrum is mainly given by electrons emitted directly from the Cor inside the droplets. For comparison, PES of gasphase Cor (black)~\cite{acocella:2016} and of solid Cor are also shown (grey curve)~\cite{deleuze:2004,yamakado:1998}. To take into account the finite resolution of our VMI spectrometer, we convolve the gasphase spectrum with a gaussian function with at full width at half maximum (FWHM) of 6\,\% of the peak energy, see the light blue line in Figure 2. This spectrum clearly resembles the one measured in He droplets in that the triplet of peaks labeled E-G is the dominant feature and peaks A-D are also visible. However, peaks A-D are significantly less pronounced in proportion to peaks E-G. In this respect, the droplet PES more closely follows the PES measured for solid Cor, in which peaks E-G are the dominant features. 


To test whether elastic scattering of the photoelectrons upon He atoms on their way out of the He droplets could account for the redistribution of peak intensity from peaks A-C to E-G we have also carried out a classical 3-D scattering simulation based on the differential electron-He scattering cross sections. A detailed description of the simulation can be found in our previous work~\cite{Shcherbinin:2018}. We find that the result of the simulation (blue curve) resembles the PES of solid Cor and to some extent our measured PES in He droplets. The effect of elastic scattering is to shift peak positions down in energy by a few 100~meV and to transfer the overall intensity towards the low-energy part of the spectrum. However, features A-C remain more pronounced in the simulation compared to the He droplet spectrum. The close resemblance with the solid-phase Cor spectrum indicates that mostly Cor clusters contribute to the He-droplet PES, as clusters are an intermediate state of matter between individual molecules and the condensed phase. Thus, while He-droplet Penning ionization electron spectra (PIES) seem to be strongly perturbed and not to contain much useful information about the dopant's electron spectrum~\cite{Shcherbinin:2018,Mandal:2020}, PES may still reveal the main features of the electron binding energies. This opens new opportunities for studying molecular complexes and nanostructures formed by He-nanodroplet aggregation at low temperature~\cite{mauracher2018cold}. 



\begin{figure}[htb]
	\center
	\includegraphics[width=1\columnwidth]{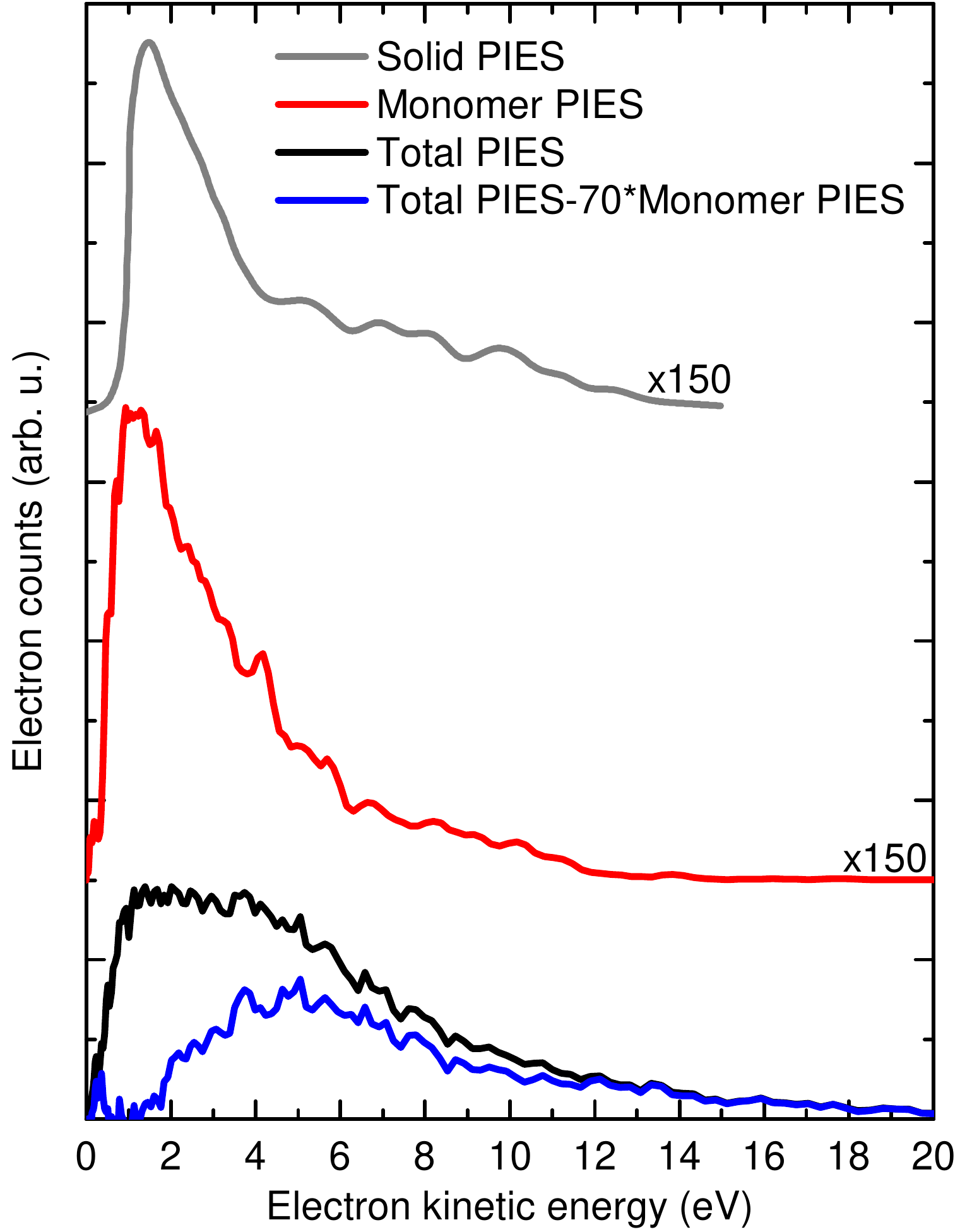}\caption{\label{fig3}
	He nanodroplet-correlated Penning ionization electron spectrum recorded in coincidence with coronene ions at $h\nu = 21.6~$eV (red line). The black line shows the total electron spectrum from Penning ionization of coronene doped He nanodroplets. The blue curve shows the difference between the total electron spectrum (black) and the spectrum measured in coincidence (red) with a scaling factor of 1/70. The grey curve is the Penning ionization spectrum of solid coronene taken from~\cite{munakata:1978}.}
\end{figure}
For comparison, we have also measured the PIES in coincidence with Cor$^+$ for Cor-doped He nanodroplets, see the red line in Fig.~3. In contrast to the PES, this spectrum is nearly structureless and peaks at low energies, around 1.2~eV. It largely resembles PIES measured for other PAH molecules doped in He nanodroplets~\cite{Shcherbinin:2018}. Note that the PIES measured for a solid Cor surface (grey curve in Fig.~3) is also dominated by a peak at 1.5~eV~\cite{munakata:1978}. However, part of the spectrum between 4 and 14~eV still reveals some structure reminiscent of the electron binding-energy spectrum. In that experiment, the contribution of both 1s2s\,$^1$S and $^3$S excited He atoms lead to the congestion of the PIES.

The signal minimum at electron energies between 0 and 1~eV in the He-droplet PIES, which was previously observed for other dopants, is likely related to the energy gap between the ionization energy of He atoms and the lower conduction-band edge of electrons in liquid He which causes low-energy electrons to be trapped and to recombine with the ion~\cite{Wang:2008,Shcherbinin:2018,mauracher2018cold}. Merely the onset of the electron signal at 14.6~eV is reminiscent of the PES when accounting for the lower photon energy  $h\nu=18.5~$eV compared to the internal energy of 1s2s\,$^1$S-excited He (20.6~eV) inducing Penning ionization. The large difference between the PES (Fig.~2) and the PIES (Fig.~3) shows that perturbations of the electron energies cannot solely be due to the interaction of the emitted electron with the He droplet which should be the same for both cases, He-droplet PES and PIES. Apparently, the Penning ionization process itself is more strongly affected by the He shell surrounding the dopant than the photoemission process. This finding should be further investigated both experimentally and theoretically.

In addition to the He-droplet PIES measured in coincidence with Cor$^+$, Fig.~3 shows the PIES measured when recording all emitted electrons (black line). To eliminate contributions from ionization of the residual gas, we subtracted the corresponding spectrum measured when the He droplet beam is blocked. The resulting electron spectrum appears to contain an additional component around 5~eV, which we tentatively inferred by subtracting the coincidence PIES weighted by factor 1/70, see the blue line. This electron component must stem from ionization events where the ion eludes its detection as it is not present in the coincidence PIES. Most likely it is due to Penning ionization of larger Cor clusters whose ions remain bound inside the He droplets, as it occurs for xenon atoms~\cite{Wang:2008,ltaief:2020direct}. Differences between the coincidence and total electron PIES of alkali metal dopants have recently revealed details about the Penning ionization process which we dicussed in the context long-range and short-range interatomic Coulombic decay~\cite{Ltaief:2019}. However, this concept does not seem to apply in the present case as neither of the PIES reproduce the expected electron binding energy spectrum.

\begin{figure}[htb]
	\center
	\includegraphics[width=1\columnwidth]{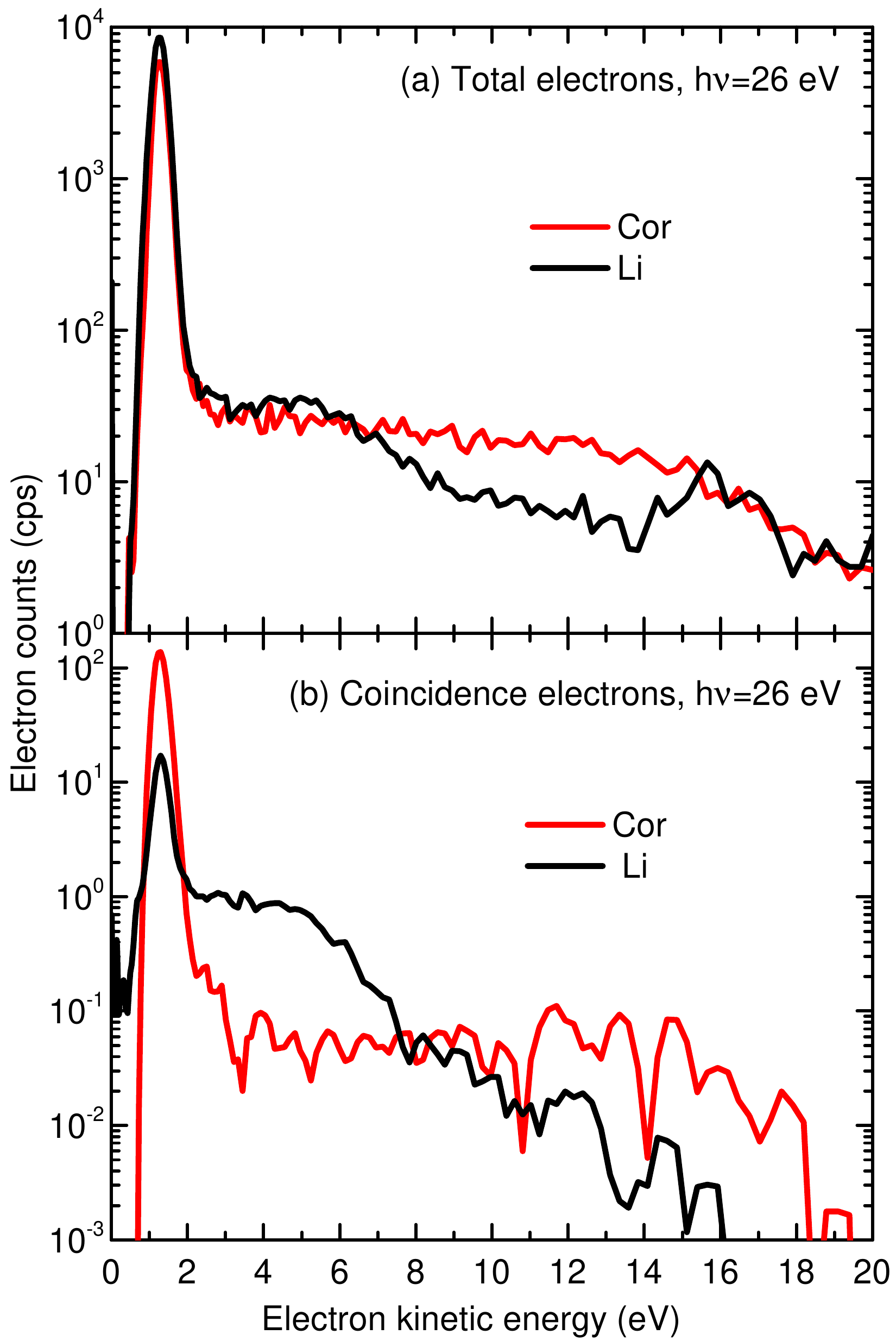}\caption{\label{fig4} Comparison between electron spectra recorded for He nanodroplets doped with coronene and with lithium atoms in the regime of droplet photoionization ($h\nu = 26~$eV). (a) Spectra of all emitted electrons. (b) Spectra of electrons detected in coincidence with coronene and lithium ions.}
\end{figure}

For completeness, we also show electron spectra recorded at $h\nu = 26$~eV, \textit{i.~e.} in the regime of direct photoionization of the He nanodroplets, see the red lines in Fig.~4. Panel a) displays the spectra of all emitted electrons and panel b) shows the electron spectra recorded in coincidence with the dopant ions. It is interesting to compare with previously measured spectra of He droplets doped with lithium (Li) atoms as the He-droplet PIES of alkali metals are very well resolved~\cite{Buchta:2013,Ltaief:2019}. The dominant feature is the He photoline at 1.4~eV, even in the spectra recorded in coincidence with the dopant ions. This indicates efficient charge-transfer ionization of the dopants. 

Surprisingly, the spectra contain a broad shoulder extending to 18~eV at a lower signal level by nearly 3 orders of magnitude. This signal appears to be due to Penning ionization as its onset roughly matches the electron energies one expects for Penning ionization, which are in the ranges 13.2-17.2~eV for Cor and 15.2-19.2~eV for Li for the case of He in the 1s2s\,$^1$S state up to high Rydberg states. The total-electron spectrum recorded for Li doping [grey line in Fig. 4 a)] even features a clear maximum around 16~eV. This shows that even at a photon energy 1.4~eV above the ionization threshold of He, a small fraction of the electrons remain bound to the droplets, partly relax to lower He excited states, and undergo Penning ionization. Previously, this effect was only observed for $h\nu\leq 25~$eV, that is below the lower conduction band edge of liquid He~\cite{Buchta:2013,Mandal:2020}. This highlights the strong coupling of quasi-free electrons to He nanodroplets and the capability of He droplets to efficiently dissipate energy by inducing electronic relaxation and Penning ionization.

\section{Conclusion}
In summary, we have presented the first one-photon extreme-ultraviolet photoelectron spectrum of a dopant molecule, coronene, embedded in He nanodroplets. Within the experimental error, the spectrum closely resembles that of solid coronene, indicating that predominantly corone clusters are photoionized. Penning ionization electron spectra can be measured with more than a factor 100 higher count rates owing to the large resonant absorption cross section of He droplets and the efficient channeling of the optical excitation to the dopant molecule. However, those electron spectra are nearly structureless, indicating strong perturbation of the Penning process by the He droplet. 

Although coronene is a favorable case due to its large photoionization cross section in a suitable range of photon energies, this work proves the principle that ultraviolet photoelectron spectroscopy (UPS) and likely x-ray photoelectron spectroscopy (XPS) of dopants in He droplets is possible. This opens the way to more systematic structural investigations of molecular complexes and nano-clusters formed in the unique ultracold environment of superfluid He nanodroplets. However, the low target density and the lack of enhancement of dopant ionization by the He droplet requires sensitive detection techniques and long acquisition times.

\section*{Acknowledgement}
M.M. and L.B.L. acknowledge financial support by Deutsche Forschungsgemeinschaft (DFG, German Research Foundation, projects MU 2347/10-1 and BE 6788/1-1) and by the Carlsberg Foundation. SRK thanks DST and MHRD, Govt of India, through the IMPRINT programmes, and the Max Planck Society. M.M. and S.R.K. gratefully acknowledge funding from the SPARC Programme, MHRD, India. The research leading to this result has been supported by the project CALIPSOplus under grant agreement 730872 from the EU Framework Programme for Research and Innovation HORIZON 2020.

%


\end{document}